# A derivation of Maxwell's equations using the Heaviside notation

**Damian P Hampshire**

**Superconductivity Group, Centre for Materials Physics,
Department of Physics, University of Durham, South Road,
Durham, DH1 3LE, United Kingdom.**

## Summary

Maxwell's four differential equations describing electromagnetism are amongst the most famous equations in science. Feynman said that they provide four of the seven fundamental laws of classical physics. In this paper, we derive Maxwell's equations using a well-established approach for deriving time-dependent differential equations from static laws. The derivation uses the standard Heaviside notation. It assumes conservation of charge and that Coulomb's law of electrostatics and Ampère's law of magnetostatics are both correct as a function of time when they are limited to describing a local system. It is analogous to deriving the differential equation of motion for sound, assuming conservation of mass and that Hooke's static law of elasticity holds for a system in local equilibrium. We demonstrate that Faraday's law can be derived without any relativistic assumptions about Lorentz invariance and discuss creation of charge.

## 1. Introduction

This paper is written in commemoration of 125 Years of Oliver Heaviside's Electromagnetic Theory: Physical and Engineering Science Papers and Historical Perspectives. Heaviside was broadly self-taught, an eccentric, and a fabulous electrical engineer. He very probably first read Maxwell's great treatise on electricity and magnetism [1], while he was in the library of the Literary and Philosophical Society of Newcastle-upon-Tyne, just up the road from here in Durham [2]. Heaviside restructured Maxwell's original twenty equations to be the four equations that we now recognise as Maxwell's equations. He called Maxwell 'heaven-sent' [3]. In every high school, good physics students can write down Newton's laws. In every University, they can write down Maxwell's equations in the mathematical form developed by Heaviside.

Axioms in mathematics play a central pedagogical role in learning and understanding this discipline. In 300 BC, Euclid wrote 'Elements', his seminal text about geometrical mathematics [4]. It included his ten axioms and the proofs of more than four hundred propositions or theorems. It has provided the template for the logical approach that students have used over the following two and a half millennia. Students demonstrate their knowledge and skill by using the axioms as starting points and derive all the consequences that follow. Axioms in science are usually taken to be generally true, or at least very widely true, and are distinguished from those more limited statements or equations that can be derived from the axioms and then used to describe a particular system or to provide results for an examination. Hence we expect the professional mathematical and scientific communities to specify the axioms of their disciplines clearly and mark the development of new knowledge by changes in axioms. There is also the expectation that the most useful axioms, as the Greek word axioma (self-evident truth or starting point) suggests, cannot be derived from other equations or laws.

Probably the most famous physics textbook of modern times is the three volume textbook "The Feynman Lectures on Physics". In it, Feynman says "we can understand the complete realm of classical physics" from just seven equations [5]. The first three equations describe forces: Newton's law of motion, Newton's law of gravity, and the force law for a charged particle moving in a magnetic and electric field. The remaining four are Maxwell's differential equations. Students of electromagnetism are introduced to Maxwell's equations and taught that they are generally true, not least because of the overwhelming body of experimental data that validates them. Not only do they describe the **_E_**-fields and **_B_**-fields from charges and currents in vacuum but by considering the charges and currents produced in materials, they describe the fields produced by all the important technologically useful materials and an enormous range of physical phenomena in the world around us. They also include a prohibition on the creation of net charge that is consistent with all experimentation to date.

In Maxwell's original work, he used a heuristic approach to derive twenty scalar equations that describe electromagnetism. He was the first to demonstrate that light is a transverse electromagnetic wave. The equations have a form that follows Newton and emphasise the electromotive force produced by electric and magnetic fields, as shown in Table 1. Heaviside took the equations, eliminated the vector and scalar potentials and developed the differential vector calculus notation necessary to write them down in the form that we currently use. Heaviside's form gives the **_E_**- and **_B_**- fields an importance beyond the forces they can produce and opens the way to describe wave and energy propagation more directly. The historical development of electromagnetism has influenced it's modern-day teaching. Undergraduate textbooks derive the electrostatic and magnetostatic differential equations mathematically from Coulomb's law and Ampere's law. However to arrive at Maxwell's time-dependent equations, students follow the heuristic approach. Most science students are then taught relativity without understanding properly the axioms of classical electromagnetism. This is pedagogically unsound because if we do not make explicit the axioms of classical electromagnetism, in the (albeit unlikely) event that there are new experiments that are not consistent with current understanding, we undermine our students ability to identify which axioms can be retained and which ones should be discarded. For example many students think that Faraday's law is axiomatic or that the postulates of relativity are required to derive Faraday's law. In this paper, we show that Faraday's law can be derived without using any of the assumptions from Einstein's theory of relativity. Indeed the derivations here beg the question as to whether Coulomb's law, Ampère's law and Faraday's law should all have the status of laws at all, given that we can derive Faraday's law from the other two.

The next section of this paper discusses the process by which static laws can be used to derive time-dependent differential equations. As an exemplar, it considers the textbook use of Hooke's static law of elasticity to derive the time-dependent differential equation that describes the propagation of sound. Section 3 uses a similar approach to derive Maxwell's equations. We apply the vector calculus approach developed by Heaviside [6] to derive all four of Maxwell's equations. Finally we speculate about possible sources of experimental evidence for the break-down of Maxwell's equations.

## 2. Deriving time-dependent differential equations from static laws

Scientists are well versed in using static laws to derive time dependent partial differential equations. To derive the time-dependent differential equation for the propagation of sound, we start with Hooke's static law of elasticity, that when used to describe static equilibrium in a gas, can be written

$$p - p(0) = \frac{B}{\rho_D(0)}(\rho_D - \rho_D(0)),  \qquad 2\text{-}1$$

where B is the bulk modulus, $p(0)$ and $\rho_D(0)$ are the initial pressure and density of the gas under test, $p$ is the applied pressure and $\rho_D$ the resultant density. Hooke's static law is then rewritten as

$$\left(\frac{\partial p}{\partial \rho_D}\right)_t = \frac{B}{\rho_D(0)}. \qquad 2\text{-}2$$

Equation 2-1 and 2-2 are quite different types of equations. Equation 2-1 relates how a change in the external pressure applied to a uniform and static gas, changes the density throughout the entire gaseous system. Equation 2-2 is a differential equation that describes how a differential pressure across an infinitesimal volume causes a differential change in density. We note that equation 2-2 is derived by considering an element in which the cause ($\partial p$) and the effect ($\partial \rho_D$) are infinitesimally close together (i.e. local). We describe the gas as being in local equilibrium, so that even though the pressure and density can vary as a function of space and time, every point throughout the system has local values of $\rho$ and $p$ related by Hooke's law. Local equilibrium also ensures that the differential of pressure with respect to space or time, is related to an equivalent differential for density, for example by an equation of the form

$$\left(\frac{\partial p}{\partial t}\right)_x = \frac{B}{\rho_D(0)}\left(\frac{\partial \rho_D}{\partial t}\right)_x. \qquad 2\text{-}3$$

However, it is important to note that strictly, it is *not possible* to derive either equation 2-2 or 2-3 from 2-1 using mathematics alone. In 2-1, $\rho$ and $p$ do not include the variables $x$ and $t$ (i.e. space and time), whereas in 2-2 and 2-3 they are functions of $x$ and $t$. Text-books do not usually emphasise that we have used our physical intuition and followed Occam's Razor [7], so that among the many possible dependencies that include the time dependence for $p$ and $\rho_D$, we have selected the one with the fewest additional assumptions.

To derive the equation that describes the propagation of sound, we then use Newton's second law of motion in the form

$$\left(\frac{\partial p}{\partial x}\right)_t = -\rho_D(0)\left(\frac{\partial u}{\partial t}\right)_x, \qquad 2\text{-}4$$

where $u$ is the velocity and $t$ is the time. Using the identity $\left(\frac{\partial p}{\partial x}\right)_t = \left(\frac{\partial p}{\partial \rho_D}\right)_t \left(\frac{\partial \rho_D}{\partial x}\right)_t$ and equation 2-2, Newton's law gives:

$$\left(\frac{\partial \rho_D}{\partial x}\right)_t = -\frac{\rho_D^2(0)}{B}\left(\frac{\partial u}{\partial t}\right)_x \qquad 2\text{-}5$$

We then use the conservation of mass:

$$\left(\frac{\partial \rho_D}{\partial t}\right)_x = -\rho_D(0)\left(\frac{\partial u}{\partial x}\right)_t \qquad 2\text{-}6$$

and partially differentiating 2-5 with respect to *t* and partially differentiating 2-6 with respect to *x* and allowing changes in the order of differentiation, we find:

$$\left(\frac{\partial^2 u}{\partial t^2}\right)_x = \frac{B}{\rho_D(0)}\left(\frac{\partial^2 u}{\partial x^2}\right)_t. \qquad 2\text{-}7$$

From 2-7, the velocity of sound, v, is given by $v^2 = B/\rho_D(0)$. The extension of Hooke's static law to the time domain allows us to describe a whole new range of phenomena associated with pressure waves (e.g. sound). However, this derivation also serves as a useful reminder of the limitations with this approach. In practice, the propagation of sound in a gas does not strictly obey 2-7, since propagation depends on how the temperature changes while the pressure is changing. Experimental results show that new physics, not found in the static measurements, is relevant in time-dependent systems (i.e. the rate of heat flow). There are other examples of systems in physics, where we start with a static law and can derive time-dependent differential equations some of which to first order do not require additional terms (e.g. in deriving dispersion relations such as the classical derivation for magnons using the Heisenberg spin Hamiltonian) and other examples where additional frictional terms are added (e.g. in describing energy loss such as in dispersive resonant polarisation in dielectrics). Hence we emphasize that the validation of any time-dependent equations is ultimately an issue for experimentation. In this paper, we follow the simple approach described in Equations 2-1 to 2-7. We postulate that if Coulomb's law of electrostatics and Ampère's law of magnetostatics are limited to describe what could be called 'local equilibrium' - a local point of observation with local charges and current densities (i.e. local cause and effect), and written in the most simple time-dependent form (invoking Occam's Razor), then the derivatives of these laws, the differential equations with respect to time and space, hold throughout the whole system. The philosophy of this approach looks to make the system sufficiently general that it includes all the components necessary to provide general differential equations.

# 3. Derivations of Maxwell's four equations

## 3.1 The Divergence of $\underline{E}$

The most simple generalization of Coulomb's law of electrostatics, to a time-dependent form where the point of observation and the charges present are local, is

$$\underline{E}(\underline{r},t) = \lim_{\eta \to 0} \frac{1}{4\pi\varepsilon_0} \int \frac{\hat{\underline{\eta}}}{\eta^2} \rho'(\underline{r}',t)\,d\tau', \qquad 3\text{-}8$$

where as shown in Fig. 1, the electric field, $\underline{E}$, at a point of observation P located at a point $\underline{r}(x,y,z)$ and time $t$, is produced by the charge densities $\rho'(\underline{r}',t)$ located at the primed points $\underline{r}'(x',y',z')$ at the same time $t$. By definition, $\underline{\eta} = \underline{r} - \underline{r}'$ and $d\tau'$ denotes integrating over the primed spatial variables of the charge densities while the unprimed spatial variables remain constant. $\rho'(\underline{r}',t)$ is the charge density at $\underline{r}'$. $\underline{E}$ is a function of the unprimed spatial variables $x$, $y$, $z$ as well as time $t$. We assume that all the charge densities are local – very close to the point of observation. Hence the partial time derivative of the $\underline{E}$-field at the point of observation is

$$\frac{\partial \underline{E}}{\partial t} = \lim_{\eta \to 0} \frac{1}{4\pi\varepsilon_0} \int \frac{\hat{\eta}}{\eta^2} \frac{\partial \rho'}{\partial t} \, d\tau', \qquad 3\text{-}9$$

where $\partial \rho'/\partial t$ is calculated at time $t$. As is standard convention, partial derivatives with respect to time are calculated assuming all spatial variables (i.e. primed and unprimed) are held constant. We state the standard definition of the del operator $\underline{\nabla}$:

$$\underline{\nabla} = \hat{i}\frac{\partial}{\partial x}\bigg)_{y,z} + \hat{j}\frac{\partial}{\partial y}\bigg)_{x,z} + \hat{k}\frac{\partial}{\partial z}\bigg)_{x,y}, \qquad 3\text{-}10$$

and note that for this operator, in addition to the unprimed spatial variables that are explicitly shown to be held constant, for each of the partial derivatives, the variable $t$, and the primed variables $x'$, $y'$ and $z'$, are also held constant. Equations 3-1 and 3-3 lead to

$$\underline{\nabla} \cdot \underline{E} = \frac{1}{4\pi\varepsilon_0} \lim_{\eta \to 0} \int \underline{\nabla} \cdot \left(\frac{\hat{\eta}}{\eta^2} \rho'(\underline{r}',t)\right) d\tau'. \qquad 3\text{-}11$$

Using the identities: $\underline{\nabla} \cdot \left(\frac{\hat{\eta}}{\eta^2} \rho'\right) = \rho' \underline{\nabla} \cdot \left(\frac{\hat{\eta}}{\eta^2}\right) + \frac{\hat{\eta}}{\eta^2} \cdot \underline{\nabla} \rho'$ and $\underline{\nabla} \cdot \frac{\hat{\eta}}{\eta^2} = 4\pi\delta^3(\underline{\eta})$ and noting that $\rho'$ only depends on primed variables and the time $t$, we obtain one of Maxwell's equations,

$$\underline{\nabla} \cdot \underline{E} = \frac{\rho}{\varepsilon_0} . \qquad 3\text{-}12$$

Equation 3-5 has the same form and uses similar mathematical identities to those used to derive the standard result from electrostatics. However in this paper, we derive it from local time-dependent equations and then assume it is one of the underlying or fundamental equations that is correct at all points in space and time. Maxwell's equations have no agreed order. We call it Maxwell's first equation.

## 3.2  The Divergence of $\underline{B}$

We use Ampère's law of magnetostatics and again invoke Occam's Razor to postulate that the local time-dependent $\underline{B}$-field at time $t$ is

$$\underline{B}(\underline{r},t) = \lim_{\eta \to 0} \frac{\mu_0}{4\pi} \int \underline{J}'(\underline{r}',t) \times \frac{\hat{\eta}}{\eta^2} d\tau', \qquad 3\text{-}13$$

where $\underline{J}'$ is only a function of the primed spatial variables and the time is $t$. Again we assume that equation 3-6 is only valid for a system where all the current densities are local to the point of observation. We can also write

$$\frac{\partial \underline{B}}{\partial t} = \lim_{\eta \to 0} \frac{\mu_0}{4\pi} \int \frac{\partial \underline{J}'}{\partial t} \times \frac{\hat{\eta}}{\eta^2} d\tau', \qquad 3\text{-}14$$

where $\partial \underline{J}'/\partial t$ is calculated at $t$. To improve brevity, we will omit including $\lim_{\eta \to 0}$ in the integral equations in this paper hereafter. Using a general vector field identity written in the form $\underline{\nabla} \cdot \left(\underline{J}' \times \frac{\hat{\eta}}{\eta^2}\right) = \frac{\hat{\eta}}{\eta^2} \cdot \left(\underline{\nabla} \times \underline{J}'\right) - \underline{J}' \cdot \left(\underline{\nabla} \times \frac{\hat{\eta}}{\eta^2}\right)$ [8], given $\underline{\nabla} \times \underline{J}' = 0$ (because $\underline{\nabla}$ is not primed but $\underline{J}'$ is primed), the divergence of 3-6 leads to

$$\underline{\nabla} \cdot \underline{B} = \frac{\mu_0}{4\pi} \int -\underline{J}' \cdot \underline{\nabla} \times \frac{\hat{\eta}}{\eta^2} d\tau'. \qquad 3\text{-}15$$

Using the vector field identity $\underline{\nabla} \times \frac{\hat{\eta}}{\eta^2} = 0$ leads to the second of Maxwell's equations:

$$\underline{\nabla} \cdot \underline{B} = 0. \qquad 3\text{-}16$$

This equation also has the same form and uses similar mathematical identities to those used to derive the standard result from magnetostatics. However, as noted above, we have derived it from local time-dependent equations and assume it is correct at all points in space and time.

### 3.3  The Curl of $\underline{B}$

#### 3.3.1 Maxwell's Displacement Current Density

Textbooks [9] show, by taking the curl of both sides of the Ampère's magnetostatic law, that

$$\underline{\nabla} \times \underline{B} = \mu_0 \underline{J}. \qquad\qquad 3\text{-}17$$

Maxwell realized that equation 3-10 cannot be generally true as a function of time, given the vector field identity $\underline{\nabla} \cdot (\underline{\nabla} \times \underline{B}) = 0$. By invoking the continuity of charge equation given by $\underline{\nabla} \cdot \underline{J} = -\partial \rho / \partial t$ and considering the partial time derivative of 3-5, he added his famous displacement current density term $\varepsilon_0\, \partial \underline{E}/\partial t$ to equation 3-10 to give the third of his equations:

$$\underline{\nabla} \times \underline{B} = \mu_0 \underline{J} + \mu_0 \varepsilon_0 \frac{\partial \underline{E}}{\partial t}. \qquad\qquad 3\text{-}18$$

Another approach used to justify the generalization from the magnetostatic equation 3-10 to the time-dependent equation 3-11 is found by considering Figure 2. A current density flows in wires to charge capacitor plates and produces a changing $\underline{E}$-field between the plates. Using Stoke's theorem, one can rewrite 3-10 in terms of the line integral of the magnetic field around the path that bounds surface A and the surface integral across surface A where

$$\int \underline{B} \cdot d\underline{l} = \mu_0 \int \underline{J} \cdot d\underline{S}, \qquad\qquad 3\text{-}19$$

$\underline{J}$ is the current density in the wire and $\underline{S}$ is the cross-sectional area of the wire. However for 3-12 to describe correctly the $\underline{B}$-field produced in Fig. 2, the line integral $\int \underline{B} \cdot d\underline{l}$ must not depend on whether we choose surface A or surface B over which to complete the surface integral. The right-hand-side of 3-12 is $\mu_0 \int \underline{J} \cdot d\underline{S}$ for surface A and zero for surface B because no current passes through surface B. To ensure that the line integral of $\underline{B}$ doesn't depend on whether surface A or surface B is chosen, and noting that the current density in the wire is given by $\underline{J} = \varepsilon_0\, \partial \underline{E}/\partial t$, where $\underline{E}$ is the field between the plates, one can add Maxwell's displacement current density term to 3-10 to produce 3-11. Maxwell's brilliant addition led to the unification of electricity and magnetism.

#### 3.3.2 Maxwell's Third Equation

In deriving Maxwell's third (and fourth) equation, we assume the system is constrained by the conservation of charge. The constraint implies

$$\frac{\partial \rho'}{\partial t} = -\underline{\nabla}' \cdot \underline{J}', \qquad\qquad 3\text{-}20$$

where $\underline{J}'$ and $\rho'$ are the current density and charge density at the point $\underline{r}'$. We have used the standard definition (cf Equation 3-3)

$$\underline{\nabla}' = \hat{\mathbf{i}} \frac{\partial}{\partial x'}\bigg)_{y',z'} + \hat{\mathbf{j}} \frac{\partial}{\partial y'}\bigg)_{x',z'} + \hat{\mathbf{k}} \frac{\partial}{\partial z'}\bigg)_{x',y'}. \qquad\qquad 3\text{-}21$$

For this operator, similarly to Equation 3-3, in addition to the primed spatial variables explicitly shown, the variable $t$, and the non-primed variables $x$, $y$ and $z$, are also held constant. Substituting 3-13 into 3-2 and then using standard vector field manipulations that include changing the order of partial derivatives and the vector field identity $\underline{\nabla}(1/\eta) = -\hat{\underline{\eta}}/\eta^2$, we find that

$$\frac{\partial \underline{E}}{\partial t} = \frac{1}{4\pi\varepsilon_0} \underline{\nabla} \int \frac{1}{\eta} \underline{\nabla}' \cdot \underline{J}' \, d\tau'. \qquad 3\text{-}22$$

Textbook vector field algebraic techniques used in magnetostatics for functions of just three spatial variables can be used to rearrange the right-hand-side of 3-15. Using the identity :

$$\underline{\nabla}' \cdot \left[\frac{1}{\eta}\underline{J}'\right] = \frac{1}{\eta}\underline{\nabla}' \cdot \underline{J}' + \underline{\nabla}'\left(\frac{1}{\eta}\right) \cdot \underline{J}'. \qquad 3\text{-}23$$

and then integrating gives:

$$\int \frac{1}{\eta} \underline{\nabla}' \cdot \underline{J}' \, d\tau' = \int \underline{\nabla}' \cdot \left[\frac{1}{\eta}\underline{J}'\right] d\tau' - \int \underline{\nabla}'\left(\frac{1}{\eta}\right) \cdot \underline{J}' \, d\tau'. \qquad 3\text{-}24$$

The second of the three integrals in 3-17 can be written as a surface integral and then set to zero, since without loss of generality we can assume $\underline{J}' = 0$ over the surface that defines $\tau'$. Using 3-17, $\underline{\nabla}'(1/\eta) = -\underline{\nabla}(1/\eta)$, 3-16 with the del operator unprimed and $\underline{\nabla} \cdot \underline{J}' = 0$, (because $\underline{\nabla}$ is not primed but $\underline{J}'$ is primed), 3-15 then becomes:

$$\frac{\partial \underline{E}}{\partial t} = -\frac{1}{4\pi\varepsilon_0} \underline{\nabla} \int \underline{\nabla}'\left(\frac{1}{\eta}\right) \cdot \underline{J}' \, d\tau' = \frac{1}{4\pi\varepsilon_0} \underline{\nabla} \int \underline{\nabla}\left(\frac{1}{\eta}\right) \cdot \underline{J}' \, d\tau' = \frac{1}{4\pi\varepsilon_0} \underline{\nabla} \int \underline{\nabla} \cdot \left(\frac{\underline{J}'}{\eta}\right) d\tau'. \qquad 3\text{-}25$$

Using the vector field identity for the curl of the curl of a vector field, $\nabla^2(1/\eta) = -4\pi\delta^3(\underline{\eta})$, the vector field identity for the curl of the product of a vector field and a scalar, $\underline{\nabla} \times \underline{J}' = 0$, and $\underline{\nabla}(1/\eta) = -\hat{\underline{\eta}}/\eta^2$ gives:

$$\frac{\partial \underline{E}}{\partial t} = \frac{1}{4\pi\varepsilon_0} \int \nabla^2\left(\frac{\underline{J}'}{\eta}\right) d\tau' + \frac{1}{4\pi\varepsilon_0} \int \underline{\nabla} \times \left(\underline{\nabla} \times \frac{\underline{J}'}{\eta}\right) d\tau' =$$

$$\frac{1}{4\pi\varepsilon_0} \int \underline{J}' \nabla^2\left(\frac{1}{\eta}\right) d\tau' - \frac{1}{4\pi\varepsilon_0} \int \underline{\nabla} \times \left(\frac{\hat{\underline{\eta}}}{\eta^2} \times \underline{J}'\right) d\tau' = -\frac{\underline{J}}{\varepsilon_0} + \frac{1}{\varepsilon_0\mu_0} \underline{\nabla} \times \underline{B} \qquad 3\text{-}26$$

Hence we find the third of Maxwell's equations:

$$\underline{\nabla} \times \underline{B} = \mu_0 \underline{J} + \mu_0\varepsilon_0 \frac{\partial \underline{E}}{\partial t}. \qquad 3\text{-}27$$

We can compare the mathematical approach we have used to derive 3-20 to Maxwell's heuristic approach. Maxwell considered a steady-state system whereas this paper considers a local system. Both have invoked Occam's Razor to generalize Coulomb's law of electrostatics to find an expression for $\partial \underline{E}/\partial t$ and are sufficiently general to find the same underlying differential equation 3-20.

## 3.4 The Curl of $\underline{E}$

### 3.4.1 Faraday's Law.

Textbooks show that Coulomb's law for electrostatics [8] leads to.

$$\underline{\nabla} \times \underline{E} = 0. \qquad 3\text{-}28$$

Equations 3-5, 3-9, 3-10 and 3-21 in time independent form, are known as the equations of electrostatics and magnetostatics. Helmholtz theorem tells us that a vector field is completely specified by knowing its divergence and curl [10]. To generalize 3-21 to include time-dependence, Maxwell used Faraday's experimental results [11]. Heaviside called Faraday 'the prince of experimentalists' [3]. Faraday found that if the magnetic field is steadily increased inside a long solenoid, there is a force on a stationary charge outside the solenoid (cf Figure 3). He measured the force on such stationary charges using loops of metallic wires (carrying unbound stationary charges) attached to voltmeters. Faraday's experiments (together with Lenz's experiments [12]) can be described mathematically as:

$$\underline{\nabla} \times \underline{E} = -k\frac{\partial \underline{B}}{\partial t}. \qquad 3\text{-}29$$

where k is a constant of proportionality. Textbooks often then assume that 3-22 is invariant under a Galilean transformation which leads to k = 1 and then further assume this result remains true even for systems where relativistic effects are important. Such assumptions are not employed in the derivation of Faraday's law below.

### 3.4.2 Maxwell's Fourth Equation

We first consider the primed partial time-derivative of equation 3-20 which is

$$\underline{\nabla}' \times \frac{\partial \underline{B}'}{\partial t} = \mu_0 \frac{\partial \underline{J}'}{\partial t} + \mu_0 \varepsilon_0 \frac{\partial^2 \underline{E}'}{\partial t^2}, \qquad \text{3-30}$$

where $\underline{B}'$, $\underline{J}'$ and $\underline{E}'$ are the magnetic field, the current density and the $\underline{E}$-field at the position $\underline{r}'$. Substituting for $\partial \underline{J}'/\partial t$ in 3-7 and using the vector field identity $\underline{\nabla}' \times (\underline{\nabla}' \times \underline{E}') - \underline{\nabla}'(\underline{\nabla}' \cdot \underline{E}') + \nabla'^2 \underline{E}' = 0$ gives

$$\frac{\partial \underline{B}}{\partial t} = \frac{1}{4\pi} \int \left\{ \left[\underline{\nabla}' \times \left(\frac{\partial \underline{B}'}{\partial t} + \underline{\nabla}' \times \underline{E}'\right)\right] + \left[\nabla'^2 \underline{E}' - \underline{\nabla}'(\underline{\nabla}' \cdot \underline{E}') - \mu_0 \varepsilon_0 \frac{\partial^2 \underline{E}'}{\partial t^2}\right] \right\} \times \frac{\hat{\eta}}{\eta^2} d\tau'. \qquad \text{3-31}$$

Using the vector field identity

$$\underline{\nabla}' \times \left(\frac{\rho' \hat{\eta}}{\eta^2}\right) = \rho' \underline{\nabla}' \times \left(\frac{\hat{\eta}}{\eta^2}\right) + \underline{\nabla}'(\rho') \times \left(\frac{\hat{\eta}}{\eta^2}\right), \qquad \text{3-32}$$

and given the curl of a radial function is zero, the second term in 3-25 is zero. Hence using the vector field identity $\underline{\nabla}(1/\eta) = -\hat{\eta}/\eta^2$ and 3-25, we can rewrite the second term in the second square bracket of 3-24 as

$$\int \left(\underline{\nabla}'(\underline{\nabla}' \cdot \underline{E}')\right) \times \frac{\hat{\eta}}{\eta^2} d\tau' = \frac{1}{\varepsilon_0} \int \left(\underline{\nabla}'(\rho')\right) \times \frac{\hat{\eta}}{\eta^2} d\tau' = \frac{1}{\varepsilon_0} \int \underline{\nabla}' \times \left(\frac{\rho'}{\eta^2} \hat{\eta}\right) d\tau'$$
$$= -\frac{1}{\varepsilon_0} \int \underline{\nabla}' \times \underline{\nabla}\left(\frac{\rho'}{\eta}\right) d\tau' = \frac{1}{\varepsilon_0} \int \underline{\nabla} \times \underline{\nabla}'\left(\frac{\rho'}{\eta}\right) d\tau'. \qquad \text{3-33}$$

Using the vector field identities $\underline{\nabla}'\left(\frac{\rho'}{\eta}\right) = \frac{1}{\eta}\underline{\nabla}'(\rho') + \rho'\underline{\nabla}'\left(\frac{1}{\eta}\right)$ and $\underline{\nabla}'(1/\eta) = \hat{\eta}/\eta^2$ gives

$$\int \underline{\nabla}'(\underline{\nabla}' \cdot \underline{E}') \times \frac{\hat{\eta}}{\eta^2} d\tau' = \frac{1}{\varepsilon_0} \underline{\nabla} \times \int \frac{1}{\eta} \underline{\nabla}'(\rho') d\tau' + \frac{1}{\varepsilon_0} \underline{\nabla} \times \int \rho' \frac{\hat{\eta}}{\eta^2} d\tau'. \qquad \text{3-34}$$

In using Coulomb's law and Ampère's law, we have ignored the internal structure of any element of charge density and current density. Hence without loss of generality, we assume that the volume occupied by every element of charge density and current density can be considered negligible and set the second integral in 3-27 to be zero. Therefore

$$\int \underline{\nabla}'(\underline{\nabla}' \cdot \underline{E}') \times \frac{\hat{\eta}}{\eta^2} d\tau' = 4\pi \, \underline{\nabla} \times \underline{E}. \qquad \text{3-35}$$

If we use the wave-equation for $\underline{E}'$ where $\nabla'^2 \underline{E}' - \mu_0 \varepsilon_0 \, \partial^2 \underline{E}'/\partial t^2 = 0$ (which is done without loss of generality) and 3-6 to cancel terms, 3-24 then gives

$$\frac{\partial \underline{B}}{\partial t} = \frac{1}{4\pi} \int \left[\underline{\nabla}' \times \left(\frac{\partial \underline{B}'}{\partial t} + \underline{\nabla}' \times \underline{E}'\right)\right] \times \frac{\hat{\eta}}{\eta^2} d\tau' - \underline{\nabla} \times \underline{E}. \qquad \text{3-36}$$

Equation 3-29 is equivalent to Faraday's law. It is the fourth of Maxwell's equations and given by:

$$\underline{\nabla} \times \underline{E} = -\frac{\partial \underline{B}}{\partial t}. \qquad \text{3-37}$$

Hence, we have shown that Coulomb's law, Ampère's law and the conservation of charge are sufficient to expect Faraday's law and that the value for the constant k in Equation 3-22 is not a matter for experimentation but is fixed to be unity. Faraday's law can be derived without any relativistic assumptions about Lorentz invariance.

## 4. Are Maxwell's Equations Universally True?

The Jefimenko equations [13] are the general solutions to Maxwell's equations where

$$\underline{E}(\underline{r},t) = \frac{1}{4\pi\varepsilon_0}\left\{\int \frac{\hat{\underline{\eta}}}{\eta^2}\rho'_r d\tau' + \int \frac{\hat{\underline{\eta}}}{\eta c}\frac{\partial \rho'_r}{\partial t} d\tau' - \int \frac{1}{\eta c^2}\frac{\partial \underline{J}'_r}{\partial t} d\tau'\right\} \qquad 4\text{-}38$$

and

$$\underline{B}(\underline{r},t) = \frac{\mu_0}{4\pi}\left\{\int \underline{J}'_r \times \frac{\hat{\underline{\eta}}}{\eta^2} d\tau' + \int \frac{\partial \underline{J}'_r}{\partial t} \times \frac{\hat{\underline{\eta}}}{\eta c} d\tau'\right\}, \qquad 4\text{-}39$$

given that $\rho'_r$ and $\underline{J}'_r$ are subject to the constraint of the conservation of charge (c.f. Equation 3-13). In the Jefimenko equations, the charge density, $\rho'_r$, and current density, $\underline{J}'_r$, are calculated at the retarded time $t_r$ where $t_r = t - \eta/c$. Jefimenko pointed out that his equations show that the fields are caused by the charge densities and current densities in the system and that when Maxwell added the displacement current density to his fourth equation, he coupled $\underline{\nabla} \times \underline{B}$ and $\partial \underline{E}/\partial t$ but did not introduce 'a cause and effect relationship' [13]. Similarly, this paper shows that Faraday's law (in the differential form given by Equation 3-30) couples $\underline{\nabla} \times \underline{E}$ and $\partial \underline{B}/\partial t$ as a result of the conservation of charge but they also should not be considered to be in a 'cause and effect relationship'. It is interesting to identify those terms in the Jefimenko general solutions that were used by Maxwell and those used in this paper to help identify Maxwell's four underlying differential equations. Maxwell used Coulomb's law and Faraday's law associated with just two of the three terms for $\underline{E}(\underline{r},t)$ in Equation 4-1. By also turning to Ampère's law and the conservation of charge, he identified both terms for $\underline{B}(\underline{r},t)$ in Equation 4-2 . He avoided the complexities of retarded time by considering a steady-state system. In this paper, the central assumptions that Coulomb's electrostatic static law and Ampere's magnetostatic law are both true in the extreme local limit as a function of time is confirmed by Jefimenko's equations (i.e. the leadings terms in Equations 4-1 and 4-2 are Equations 3-1 and 3-6 respectively). The complexity of retarded time is avoided by considering a local system where $t_r \approx t$. The $\underline{E}$-fields and $\underline{B}$-fields are coupled with the conservation of charge.

The experimental evidence for Maxwell's equations is overwhelming. Furthermore as the gateway to Einstein's theory of relativity [14], which in itself also brings its own compelling experimental evidence, any speculation about charge creation or the breakdown of Maxwell's equations, is very probably destined to be fruitless. However, because we have derived Maxwell's equations using the conservation of charge as a constraint, we complete the paper by considering what would happen if this constraint did not always apply, or more precisely, where might we look for the break-down of Maxwell's equations. We suggest that the essence of an entity that has been *created* is that there should be no experimental methods that can determine the properties of the created entity prior to creation. The probability of the entity's existence can be considered to increase from zero to one. Looking for events described by this language of probability naturally points us towards quantum mechanics. Given that quantum mechanics has been tested to exquisite accuracy and that all known interactions conserve charge, it becomes a remote possibility at best, that we can find charge creation. Alternative tests of Maxwell's equations include looking for the creation of current density, or electric and magnetic waves that do not obey Maxwell's equations. Our best chances are to seek out events that are so difficult to produce that they have not been extensively interrogated experimentally, and hence may offer something completely unexpected. We suggest investigating an Einstein-Podolsky-Rosen (EPR) experiment [15]. Typically an entangled electron-positron pair is mixed and prepared as a superposition of states with equal and opposite magnetic moments (or spins). The charges are separated and the magnetic moment or spin of the electron is measured. The well-known instantaneous collapse of the wavefunction occurs so that the positron ends up with the opposite magnetic moment (or spin) to the electron. The appearance of the moment of the positron is triggered by entirely quantum mechanical effects - no direct electromagnetic communication occurs between the electron and positron. Indeed one can think about the two charges as a single entity. However, one can argue that we do not really know how the information that leads to the positron producing a magnetic moment of opposite sign is instantaneously received - beyond asserting it is part of the fabric of quantum mechanics, or part of the nature of a macroscopic wavefunction. We suggest that while the moment of the positron is being *created* (rather than excited), the production of the $\underline{B}$-field associated with its magnetic moment may not be coupled to the production of an $\underline{E}$-field. So one could measure $(\partial \underline{E}/\partial t)_r$ and $(\partial \underline{B}/\partial t)_r$ in the wavefront of the positron, hoping to find $\underline{B}$–fields with $\underline{E}$–fields that are inconsistent with Maxwell's equations.

## 5. Conclusions and final comments

Although Euclid's choices of starting points or axioms for geometrical mathematics seem obvious, Feynman has reminded us that they are not a unique set. We can use Euclid's axioms to derive Pythagoras's theorem or we can take Pythagoras's theorem as an axiom and drop one of Euclid's axioms [16]. Hence we have the paradox that although Euclid's choices have remained generally accepted for centuries because they are closest to being self-evident truths, none of them are indispensable. The choice of axioms ultimately includes some of the 'beauty is truth, truth beauty' [17] sentiment. At the moment, most of the scientific community uses Feynman's seven equations of classical physics, including Maxwell's equations, as axioms. If we discover charge creation, then treating Maxwell's equations as axioms would become untenable. In this paper we have shown that Maxwell's equations can be justified using a mathematical derivation that follows from Coulomb's law, Ampere's law and the conservation of charge. Therefore, as with other differential equations in physics, in the unlikely event that Maxwell's equations are not true under all circumstances, we can discuss how the equations are derived and make other choices of axioms.

## Acknowledgments

I very warmly thank those with whom I have discussed the nature of axioms and truth over the years, most obviously with my wonderful wife and children, Amanda, Emily, Peter, Alexander and Michael, as well as with Andrew Davis, Douglas Halliday, Ifan Hughes, Martin McGovern, Kozo Osamura and Mark J Raine. Of course, none of them are responsible for any non-truths that may have slipped into this paper. The data are available at: http://dx.doi.org/10.15128/gh93gz492 and associated materials are on the Durham Research Online website: http://dro.dur.ac.uk/. This work is funded by EPSRC grants EP/K502832/1 for the Durham Energy Institute and grant EP/L01663X/1 for the Fusion Energy Doctoral Training Network.

## References.

**Tables**

| | | | |
|---|---|---|---|
| $\underline{J}_{Total} = \underline{J}_{free} + \dfrac{\partial \underline{D}}{\partial t}$ | (A) | $\underline{\xi} = \dfrac{\underline{D}}{\varepsilon_r \varepsilon_0}$ | (E) |
| $\mu_r \mu_0 \underline{H} = \underline{\nabla} \times \underline{A}$ | (B) | $\underline{\xi} = \rho_N \underline{J}_{free}$ | (F) |
| $\underline{\nabla} \times \underline{H} = \underline{J}_{Total}$ | (C) | $\underline{\nabla} \cdot \underline{D} = \dfrac{\rho_{free}}{\varepsilon_0}$ | (G) |
| $\underline{\xi} = -\underline{\nabla} V - \dfrac{\partial \underline{A}}{\partial t} + \underline{v} \times \mu_r \mu_0 \underline{H}$ | (D) | $\dfrac{\partial \rho_{free}}{\partial t} = -\underline{\nabla} \cdot \underline{J}_{free}$ | (H) |

Table 1: Maxwell's 20 (scalar) equations in modern form, labelled with his original lettering notation (A) – (H) [18]. The first 18 of his equations, (A) to (F), are given here as six vector equations using Heaviside's curl notation. There are also two scalar equations, (G) and (H). We have avoided Maxwell's use of 'electromotive force' and 'actual electromotive force' and taken $\underline{\xi}$ as the electromotive force per unit charge. Also, $\rho_N$ is the resistivity, $\varepsilon_r$ is relative permittivity and $\mu_r$ is relative permeability. Standard symbols have their usual meanings [19]. Equations (A) – (D) and (G) include what are now known as Maxwell's 4 equations together with the expression for the Lorentz force. The subscript 'free' that is used now for charge densities and current densities that can travel over macroscopic distances, was called 'true conduction' by Maxwell. He also considered the magnetic vector potential $\underline{A}$ in terms of the electromagnetic momentum per unit charge.

**Figures**

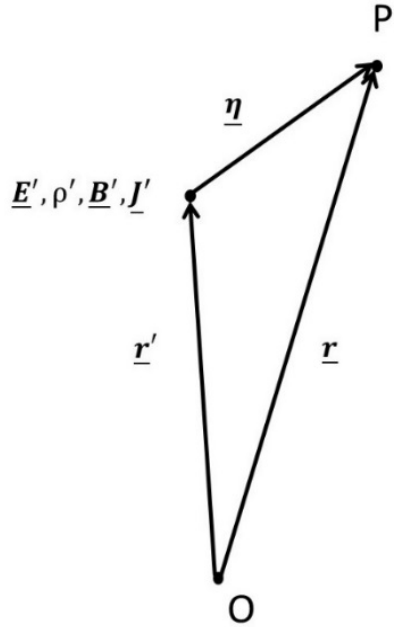

Fig 1: A coordinate system in which charge densities and current densities are observed. O is the origin. P is the point of observation. Charge densities and current densities are displaced from the origin at points $\underline{r}'$. The vector separation between the charge density $\rho'(\underline{r}')$ or current density $\underline{J}'(\underline{r}', t)$ and the observation point is given by $\underline{\eta} = \underline{r} - \underline{r}'$. The electric and magnetic fields at the primed locations are $\underline{E}'$ and $\underline{B}'$ respectively. For the special case of the electric field, charge density, magnetic field and current density at the point of observation we use unprimed values $\underline{E}$, $\rho$, $\underline{B}$ and $\underline{J}$ respectively.

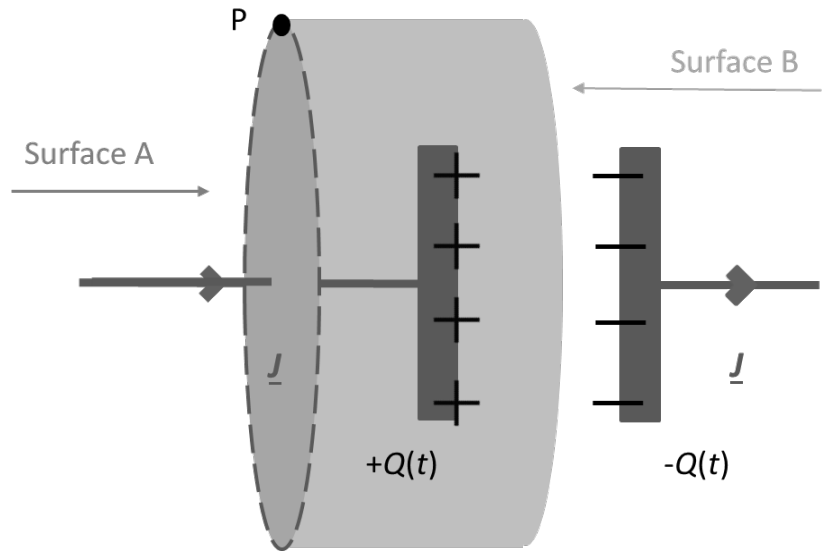

Fig 2: Straight wires carrying a constant current density (*J*) and charging two capacitor plates. Surface A is the flat circular surface bounded by the dotted ring path on which the point P is located and through which the current passes. Surface B is bounded by the same dotted ring path but passes through the capacitor plates so no current passes through it. The magnitude of the charge on each plate increases with time (*t*) and has magnitude *Q*(*t*).

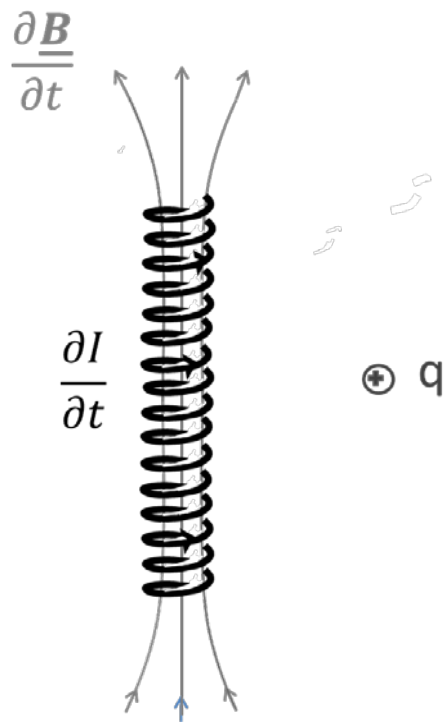

Fig 3: A stationary positive charge (q) outside a long solenoid. The current flowing in the solenoid is being increased (i.e. $\partial I/\partial t > 0$) as is the magnetic field inside the solenoid (i.e. $\partial \underline{B}/\partial t > 0$).